\documentclass[aps,pra,twocolumn,showpacs,amsmath,
amssymb,amsfonts,superscriptaddress]{revtex4}

\usepackage{graphics}
\usepackage{epsfig}

\newcommand{\beq}{\begin{equation}}
\newcommand{\eeq}{\end{equation}}
\newcommand{\beqa}{\begin{eqnarray}}
\newcommand{\eeqa}{\end{eqnarray}}

\begin{document}

\title{Quantum Gaussian Channels with Additive Correlated Classical Noise}
\author{Giovanna Ruggeri}
\email{ruggeri@le.infn.it}
\affiliation{Dipartimento di Fisica, 
Universit\`a di Lecce, I-73100 Lecce, Italy}
\author{Stefano Mancini}
\email{stefano.mancini@unicam.it}
\affiliation{Dipartimento di Fisica, Universit\`{a} di Camerino, 
I-62032 Camerino, Italy}

\begin{abstract}
We provide a model to study memory effects in quantum Gaussian channels with additive classical noise over an arbitrary number of uses. The correlation among different uses is introduced by contiguous two-mode interactions. Numerical results for few modes are presented. They confirm the possibility to enhance the classical information rate with the aid of entangled inputs, and show a likely asymptotic behavior that should  lead to the full capacity of the channel.
\end{abstract}


\maketitle

\section{Introduction}

A memoryless quantum communication channel makes the fundamental assumption that the noise between consecutive uses of the channel is independent. In many real-world applications this assumption may be good, but for many others the noise may be strongly correlated between uses of the channel. Such a possibility has been recently put forward in channels with continuous alphabet, specifically lossy bosonic channels \cite{GM05}.
The main motivation that has led to investigate memory effects in such channels
has been the possibility to enhance their classical capacity by means of entangled inputs \cite{Rug05}.

Lossy bosonic channels belong to the class of Gaussian channels which play a prominent role because it might be simple enough to be analytically tractable, thus providing a fertile testing ground for the general theory of quantum capacities \cite{Hol99}.
Another example of Gaussian channel is provided by a channel that adds thermal noise to the input.
For that channel it has been recently proved that entangled inputs enhance the 2-shot classical capacity in presence of correlated noise \cite{Cerf04}. 

Here we present an extension of the model used in Ref.\cite{Cerf04}, that can be employed for an arbitrary number of channel uses.
 The memory effect is modeled by assuming that the noise affecting subsequent uses of the channel follows a Gaussian distribution with correlation introduced by contiguous two-mode interactions. 
 We show that if the memory is non-zero and the input energy constraint is satisfied, the channel transmission rate can be improved by entangled input states instead of product states.
Moreover, we notice the appearance of an asymptotic value of the transmission rate after many uses of the channel that should lead to the full capacity of the channel.

The paper is organized as follows. In Sect.II and III, we review the basic notions about bosonic channels and Gaussian channels with additive classical noise. In Sect.IV we describe the action of the memory. Finally, in Sect.V we analyze the transmission rate of this channel and discuss our results.  


\section{Bosonic channels}

A bosonic channel is any completely positive and trace preserving map acting on the state of a mode of electromagnetic field. On multiple uses of that channel one has to account for many of such modes.
Let us consider $n$ of them associated with  $n$ pairs of annihilation and creation operators $a_{j}$ and $a_{j}^{\dagger}$ (satisfying the canonical commutation relation), or equivalently of quadrature components $q_{j}\equiv(a_j+a_j^{\dag})/\sqrt{2}$ and $p_{j}\equiv -i(a_j-a_j^{\dag})/\sqrt{2}$.
 Ordering these in a column vector 
\begin{equation}
{\bf r}=\left[ q_{1},\cdots, q_{n},p_{1},\cdots,p_{n}\right] ^{T},
\label{er}
\end{equation}
we may define the mean vector 
${\bf m}$ and the covariance matrix $\mathcal{V}$ of an $n$-mode state $\rho$ as 
\begin{equation}
{\bf m}=\mbox{Tr}\left(\rho{\bf r}\right),
\label{em}
\end{equation}
\begin{equation}
\mathcal{V}=\mbox{Tr}\left[\left( {\bf r}-{\bf m}\right) \rho \left( {\bf r}-{\bf m}\right)^{T}\right]-\frac{1}{2}\mathcal{J},
\label{calV}
\end{equation} 
where $\mathcal{J}=i\left( 
\begin{array}{cc}
0 & \mathcal{I} \\ 
-\mathcal{I} & 0
\end{array}
\right) $ is the symplectic matrix, with $\mathcal{I}$ the $n\times n$ identity
matrix.

If the state $\rho$ is Gaussian is completely characterized by the mean vector 
${\bf m}$ and the covariance matrix $\mathcal{V}$. Moreover, 
its von Neumann entropy is given by \cite{Hol99}
\begin{equation}
S\left( \rho\right)
=\sum_{j=1}^{n}g\left( \left| \lambda _{j}\right| -\frac{1}{2}\right) ,
\label{Srho}
\end{equation} 
where $\pm \lambda _{j}$ are the symplectic eigenvalues of the covariance
matrix $\mathcal{V} $, that is the solutions of
the equation 
\begin{equation}
\mbox{det}\left[ \mathcal{V}-\lambda \mathcal{J}\right] =0,
\label{char}
\end{equation}  
and
 \begin{equation}
g\left( x\right) =\left\{ 
\begin{array}{lr}
\left( x+1\right) \log _{2}\left( x+1\right) -x\log _{2}x & x>0 \\ 
0 & x=0
\end{array}
\right. 
\label{gx}
\end{equation} 
 is the entropy of a thermal state with a mean photon number $x$. 

Let us now consider the classical use of a bosonic channel
described by a completely positive and trace preserving map $T$:
\begin{equation}
\rho\mapsto T[\rho],
\label{mapT}
\end{equation}
Then, quantum states carry the values of a random classical variable. 
Quite generally we could think to map phase space points into quantum states, 
hence for each mode we consider two real random values (or one complex).
We then label an input state over $n$ modes (uses) as
 $\rho_{\mbox{\boldmath$\alpha$}}$ with $\mbox{\boldmath$\alpha$}\in \mathbb{C}^n$.

The coding theorem for quantum channels asserts that the classical
capacity of the bosonic channel $T$ is given by 
\begin{eqnarray}
C\left( T\right) &=&\lim_{n\to\infty}\frac{1}{n}\max 
\left[ S\left( \int P({d^{2n}\mbox{\boldmath$\alpha$}})T\left[ \rho _{\mbox{\boldmath$\alpha$}}\right] \right)\right.
\nonumber\\
&&\left.\quad\qquad
-\int P({d^{2n}\mbox{\boldmath$\alpha$}})S\left( T\left[ \rho _{\mbox{\boldmath$\alpha$}}\right] \right) \right] ,
\label{Cap}
\end{eqnarray}
where $S$ denotes the von
Neumann entropy and $P({d^{2n}\mbox{\boldmath$\alpha$}})$ is a probability measure.
 
In Eq.(\ref{Cap}),
the maximum is taken over all probability measures $\left\{
P({d^{2n}\mbox{\boldmath$\alpha$}})\right\} $ and collections of density operators $\left\{ \rho
_{\mbox{\boldmath$\alpha$}}\right\} $ satisfying the energy constraint 
\begin{equation}
\frac{1}{n}\int P(d^{2n}\mbox{\boldmath$\alpha$})\mbox{Tr}\left( 
\rho _{\mbox{\boldmath$\alpha$}} \sum_{j=1}^{n}
a_{j}^{\dagger}a_{j}\right) \leq \overline{n},
\label{bound}
\end{equation}
with \ $\overline{n}$ the maximum
mean photon number per mode at the input of the channel.


\section{Gaussian channels with additive classical noise}

Consider now a specific bosonic channel $T$
acting as follows 
\begin{eqnarray}
T\left[\rho_{\mbox{\boldmath$\alpha$}}\right]=\int d^{2n} \mbox{\boldmath$\beta$}\ Q( \mbox{\boldmath$\beta$})\ D( \mbox{\boldmath$\beta$})\rho_{\mbox{\boldmath$\alpha$}}^{in} D^\dagger( \mbox{\boldmath$\beta$}),
\label{outexp}
\end{eqnarray}
where $D(\mbox{\boldmath$\beta$})\equiv D(\beta_1)D(\beta_2)\ldots D(\beta_n)$ with
$D(\beta_j)$ denoting the displacement operator on the $j$th mode,  $\beta_j\in\mathbb{C}$.
This is known as \emph{Gaussian channel with additive classical noise} once the kernel is Gaussian, e.g.
\begin{equation}
Q(\mbox{\boldmath$\beta$})=\prod_{k=1}^{n}\frac{1}{\pi N} \, e^{-\frac{|\beta_k|^2}{N}}.
\label{Qbe}
\end{equation}
In such a case the channel randomly displaces each input state
according to a Gaussian distribution, which results 
in a thermal state ($N$ is the variance of the added noise
on the quadrature components $q_j$ and $p_j$, or, equivalently
the number of thermal photons added per mode by the channel). 

Since for Gaussian channels it is conjectured that Gaussian inputs allow to achieve the capacity \cite{Hol99}, let us consider to encode $\mbox{\boldmath$\alpha$}$ into coherent input states
\begin{equation}
\rho_{\mbox{\boldmath$\alpha$}}^{in}= D(\mbox{\boldmath$\alpha$})|{\bf{0}}\rangle
\langle{\bf{0}}|D^{\dag}(\mbox{\boldmath$\alpha$}),
\label{rhoin}
\end{equation}
where $D(\mbox{\boldmath$\alpha$})\equiv D(\alpha_1)D(\alpha_2)\ldots D(\alpha_n)$ 
and
$|{\bf 0}\rangle\equiv|0\rangle|0\rangle\ldots|0\rangle$ are the vacuum states of the $n$ modes. 
Moreover, suppose that the state (\ref{rhoin}) is drawn with probability 
\begin{equation}
P(\mbox{\boldmath$\alpha$})=\prod_{k=1}^{n}\frac{1}{\pi \overline{n}} \, e^{-\frac{|\alpha_k|^2}{\overline{n}}},
\label{Pal}
\end{equation}
then the output states and their average read
\begin{eqnarray}
\rho_{\mbox{\boldmath$\alpha$}}^{out}&=&T\left[\rho_{\mbox{\boldmath$\alpha$}}^{in}\right],\\
\overline{\rho}^{out}&=&\int d^{2n} \mbox{\boldmath$\alpha$}\ P( \mbox{\boldmath$\alpha$})
\rho_{\mbox{\boldmath$\alpha$}}^{out}.
\label{outave}
\end{eqnarray}

The Gaussian map effected by the Gaussian channel (\ref{outexp}),  
in case of Gaussian input state, can be solely characterized by the covariance matrices. that is
\begin{equation}
\mathcal{V}^{in} \mapsto \mathcal{V}^{in} + {\rm diag}(N,N,\ldots,N),
\label{mapV}
\end{equation}
where $\mathcal{V}^{in} $ is the input covariance matrix. For the states of Eq.(\ref{rhoin}), it is 
\begin{equation}
\mathcal{V}^{in} = \frac{1}{2}{\rm diag}(1,1,\ldots,1).
\label{calVind}
\end{equation}

We could also consider the possibility of entangled input states.
These can be accounted for by the transformation $\rho_{\mbox{\boldmath$\alpha$}}^{in}\rightarrow \Sigma(\mathcal{R})\rho_{\mbox{\boldmath$\alpha$}}^{in}\Sigma^{\dag}(\mathcal{R})$ that uses the multimode squeezing operator \cite{Lo93}
\begin{equation}
\Sigma\left( \mathcal{R} \right) =\exp \left[ \frac{1}{2}\sum_{jk=1,j \neq k}^{n} \left(
\mathcal{R}_{jk}a_{j}^{\dagger}a_{k}^{\dagger}-\mathcal{R}_{jk}^{\ast }a_{j}a_{k}\right) \right], 
\label{SigR}
\end{equation}
with $\mathcal{R}$ a symmetric $n\times x$ matrix. 
Thus, the input 
covariance matrix results
\begin{equation}
\mathcal{V}^{in}=
\frac{1}{2}\left( 
\begin{array}{cc}
\exp \left( \mathcal{R}\right)  & 0 \\ 
0 & \exp \left( -\mathcal{R}\right) 
\end{array}
\right) .
\label{calVin}
\end{equation}
Let $\overline{n}_r$ be the average number per mode of squeezed photons added at input, i.e.
\begin{equation}
\overline{n}_r=\frac{1}{n}\sum_{j=1}^{n}\langle{\bf 0}|\Sigma(\mathcal{R})
a_j^{\dag}a_j\Sigma^{\dag}(\mathcal{R})|{\bf 0}\rangle.
\end{equation}
Then, due to the bound (\ref{bound}) the effective number of  photons to input on each mode would be $(\overline{n}-\overline{n}_r)$.


\section{The memory model}

Now, let us consider memory effects in the Gaussian channel with additive classical noise.
The output state (\ref{outexp}) can be seen as the convolution of the input Gaussian state (\ref{rhoin}) with a Gaussian thermal state having average number of thermal photons $N$ per mode. To introduce correlations in the latter we use arguments from multimode squeezing.
It is physically reasonable that  only contiguous uses of the channel would be strongly correlated, thus we introduce the following symmetric matrix
 \begin{equation}
\mathcal{S}=-s\left( 
\begin{array}{ccccc}
0 & 1 & \cdots  & \cdots  & 0 \\ 
1 & 0 & 1 & \cdots  & \cdots  \\ 
\cdots  & 1 & 0 & 1 & \cdots  \\ 
\cdots  & \cdots  & 1 & 0 & 1 \\ 
0 & \cdots  & \cdots  & 1 & 0
\end{array}
\right), 
\label{calS}
\end{equation}
 where $s$ is the parameter measuring the degree of memory.
Then, we take the added noise as
characterized by the covariance matrix 
\begin{equation}
\mathcal{V}^{N}=\mathcal{V}_{1}^{N}+\epsilon \mathcal{V} _{2}^{N},
\label{calVN}
\end{equation}
 where $\mathcal{V}_1^N$ is the diagonal matrix
 \begin{equation}
[\mathcal{V}_1^N]_{jj}= N-\epsilon[\mathcal{V}_2^N]_{jj},
 \end{equation}
 and
 \begin{equation}
 \mathcal{V}^{N}_2=
\frac{1}{2}\left( 
\begin{array}{cc}
\exp \left( \mathcal{S}\right)  & 0 \\ 
0 & \exp \left( -\mathcal{S}\right) 
\end{array}
\right) .\label{calVN2}
\end{equation}
In order to realize a physical transformation, the channel must have $\mathcal{V}^N\ge 0$. This is clearly satisfied if $\mathcal{V}_1^N\ge 0$ and $\mathcal{V}_2^N\ge 0$. The first condition determines the range of allowed values for the memory parameter $s$, while the second condition is always satisfied. 
The presence of the parameter $\epsilon$ in Eq.(\ref{calVN}) only deserves as mathematical trick to guarantee the positivity of $\mathcal{V}^N$ even in the case of $N< 1/2$. It is 
\begin{eqnarray}
\epsilon=1&\quad& N\ge\frac{1}{2}\label{ep1}\\
0\le\epsilon\le 2N &\quad& N<\frac{1}{2}\label{ep2}
\end{eqnarray}
Notice that the noise correlations so introduced are classical and for $s\to 0$ in Eq.(\ref{calVN}) we recover the memoryless case.

\begin{figure}
\epsfxsize=6.5cm\epsfbox{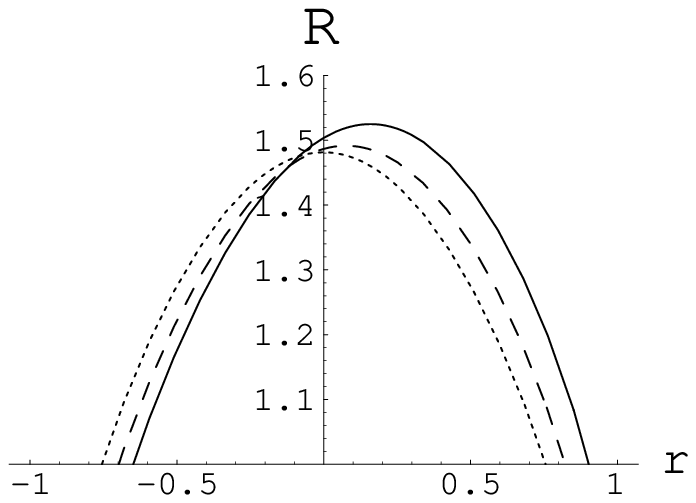}\vskip 0.2cm
\epsfxsize=6.5cm\epsfbox{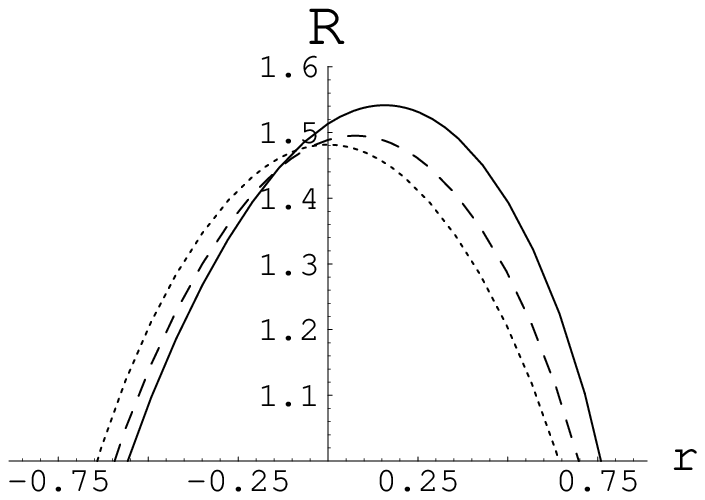}\vskip 0.2cm
\epsfxsize=6.5cm\epsfbox{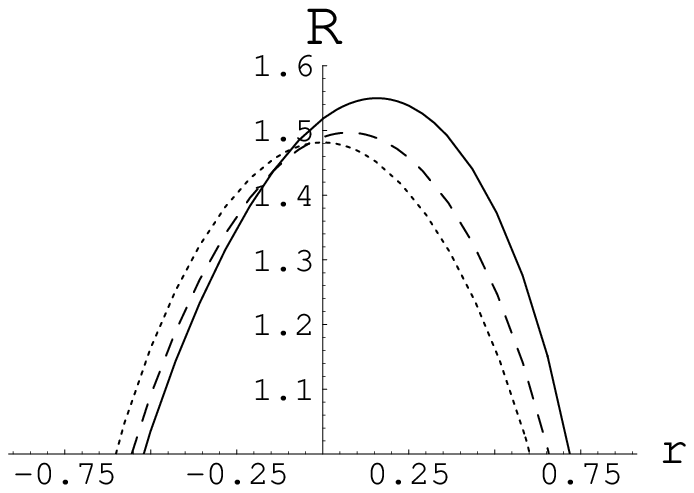}\vskip 0.2cm
\epsfxsize=6.5cm\epsfbox{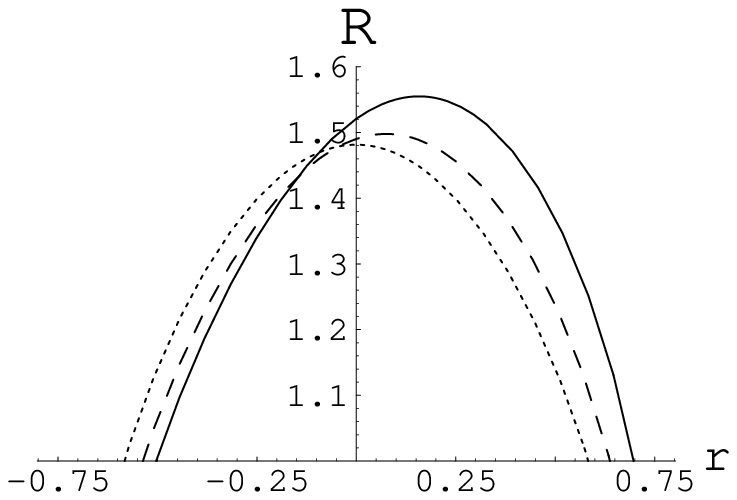}\vskip 0.2cm
\caption{Transmission rate $R$ (maximized over the correlation coefficient $y$) as a function of the input entanglement parameter $r$ for a channel with a degree of memory $s=0$ (dotted lines), $s=0.1$ (dashed lines), $s=0.2$ (solid lines). The number of uses is from top to bottom $n=2,3,4,5$. The input mean number of photons is $\overline{n}=2$, while the number of added photons is $N=\frac{2}{3}$.}
\label{fig1}
\end{figure}

For the entangled inputs we consider in Eq.(\ref{SigR}) the symmetric matrix
 \begin{equation}
\mathcal{R}=-r\left( 
\begin{array}{ccccc}
0 & 1 & \cdots  & \cdots  & 0 \\ 
1 & 0 & 1 & \cdots  & \cdots  \\ 
\cdots  & 1 & 0 & 1 & \cdots  \\ 
\cdots  & \cdots  & 1 & 0 & 1 \\ 
0 & \cdots  & \cdots  & 1 & 0
\end{array}
\right), 
\label{calR}
\end{equation}
with $r$ is the input entanglement parameter.

At the output of the channel, we get states with the covariance matrix 
\begin{equation}
\mathcal{V}^{out}=\mathcal{V}^{in}+\mathcal{V}^{N}.
\label{calVout}
\end{equation}
In turns, the covariance matrix associated with the mixture of the
output states (\ref{outave}) would be
\begin{equation}
\overline{\mathcal{V}}^{out}=\mathcal{V}^{out}+\mathcal{K}.
\label{calVoutave}
\end{equation}
Here
 \begin{equation}
\mathcal{K}=\mathcal{K}_{1}+\theta \mathcal{K}_{2}.
\label{calK}
\end{equation}
with $\mathcal{K}_1$ a diagonal matrix
\begin{equation}
[\mathcal{K}_1]_{jj}= (\overline{n}-\overline{n}_r)-\theta[\mathcal{K}_2]_{jj},
\label{calK1}
\end{equation}
and $\mathcal{K}_{2}$ being of the
same form of $\mathcal{V} _{2}^{N}$
with $s\rightarrow -y$.
The coefficient $y$ accounts for possible classical input correlations.

Also in this case we must have $\mathcal{K}\ge 0$. This is clearly satisfied if $\mathcal{K}_1\ge 0$ and $\mathcal{K}_2\ge 0$. The first condition determines the range of allowed values for the parameter $y$, while the second condition is always satisfied. 
Likewise the parameter $\epsilon$ in Eq.(\ref{calVN}), we have introduced in Eq.(\ref{calK}) a  parameter $\theta$ that only deserves to guarantee the positivity of $\mathcal{K}$ when $(\overline{n}-\overline{n}_r)< 1/2$. It is
\begin{eqnarray}
\theta=1&\quad& (\overline{n}-\overline{n}_r)\ge\frac{1}{2}\label{th1}\\
0\le\theta\le 2(\overline{n}-\overline{n}_r) &\quad& (\overline{n}-\overline{n}_r)<\frac{1}{2}.
\label{th2}
\end{eqnarray}

According to Eq.(\ref{Cap}), the capacity is given by the asymptotical behavior ($n\to\infty$) of the maximum of transmission rate
 \begin{equation}
R = \frac{1}{n}\left[S\left(\overline{\rho}^{out}\right)-S\left(\rho^{out}\right)\right].
\label{Rate1}
\end{equation}
Since $\overline{\rho}^{out}$ and ${\rho}^{out}$ are both Gaussian states,
following Eq.(\ref{Srho}) we have
\begin{equation}
R(r,y)=\frac{1
}{n}\sum_{i=1}^{n}\left[ g\left( \left| \overline{\lambda _{j}}^{out}\right| -
\frac{1}{2}\right) -g\left( \left| \lambda _{j}^{out}\right| -\frac{1}{2}
\right) \right] .\label{Rate2}
\end{equation}
Thus, in order to evaluate the transmission rate, we need
to compute the $n$ symplectic values  $\overline{
\lambda _{j}}^{out}$, $\lambda _{j}^{out}$ of  $\overline{\mathcal{V}}^{out}$, $\mathcal{V}^{out}$ respectively.

\section{Results and Conclusions}

The transmission rate of Eq.(\ref{Rate2}) has been evaluated numerically restricting to some cases where $\epsilon=\theta=1$.

From Fig.\ref{fig1}, we notice that when $s>0$, the optimized rate $R$ over $y$ increases with
the degree of entanglement $r$ and attains a maximum at some value $r^{\ast}>0$, 
so that the maximum is achieved by entangled input states and is greater than that at $s=r=0$.

\begin{figure}
\epsfxsize=6.5cm\epsfbox{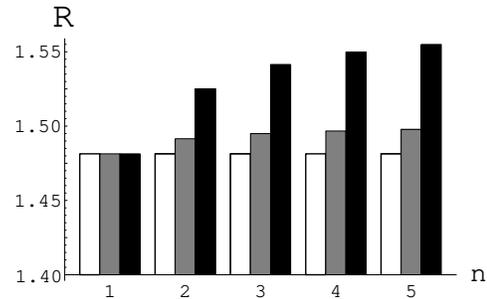}\vskip 0.2cm
\caption{The maximum of transmission rate $R$ (over the input entanglement $r$ and correlation coefficient $y$) as a function of the number of channel uses $n$ for a channel with a degree of memory $s=0$ (white bars), $s=0.1$ (grey bars), $s=0.2$ (black bars). The input mean number of photons is $\overline{n}=2$, while the number of added photons is $N=\frac{2}{3}$.}
\label{fig2}
\end{figure}

Moreover, maximizing $R$ with respect to both $r$ and $y$ allows 
to find the behavior of the transmission rate versus the number of channel uses $n$ (see Fig.\ref{fig2}). 
Here we have considered $s=0, 0.1, 0.2$, white, grey and black bars respectively. 
Unfortunately the evaluation of $R$ with larger $n$ requires a lot of numerical resources.
Nonetheless, from Fig.\ref{fig2}, we can grasp that the transmission rate increases with the number of channel uses $n$, achieving an asymptotic value depending on $s$. 
 
 This can also be understood by considering that the rate increases with noise correlations.
 Initially, they rapidly grow till $n$ reaches the relevant number of modes that effectively interact. 
 In fact, each mode effectively interacts with a limited number of other modes which would be determined (besides the form of $\mathcal{S}$) by the value of $s$ (the smaller is $s$ the smaller is the number of modes that effectively interact) . Thus the number of uses after which $R$ becomes almost constant can be thought as the range of memory effects.
 
 In conclusion, we have provided a model to describe memory effects in a Gaussian channel with additive classical noise for an arbitrary number of uses. The model reduces to that of Ref.\cite{Cerf04} for only two uses.
 From our numerical results we guess the existence of an asymptotic value of the transmission rate after many uses of the channel that should lead to the full capacity of the channel.
 Hence, the 2-shot capacity would not suffice to characterize the channel performances, unless the memory is very weak. Quite generally the minimum number of uses to consider would depend on the strength of the memory, i.e. the correlation length of the noise. 
 
Of course the results found are conditioned to to the validity of  the conjecture that Gaussian inputs allow to achieve the capacity of a Gaussian channel \cite{Hol99}, however, the presented model for memory effects is always valid and could be extended to include attenuation/amplification besides additive noise.

\section*{Acknowledgements}
\noindent
The work of S.M. has been supported by the European Commission under the Integrated Projects "QAP" and  "SCALA".

\end{document}